\documentclass[aps,showpacs]{revtex4}
\usepackage{amsmath}
\usepackage{graphicx}

\begin{document}

\title{Synchrotron radiation in Myers-Pospelov effective electrodynamics}
\author{R. Montemayor}
\affiliation{Instituto Balseiro and CAB, Universidad Nacional de Cuyo and CNEA, 8400
Bariloche, Argentina}
\author{Luis F. Urrutia}
\affiliation{Instituto de Ciencias Nucleares, Universidad Nacional Aut\'{o}noma de M\'{e}%
xico, A.P. 70-543, M\'{e}xico D.F. 04510, M\'{e}xico}

\begin{abstract}
In the framework of the classical effective Lorentz invariance violating
(LIV) model of Myers-Pospelov, we present a complete calculation of the
synchrotron radiation produced by a circularly moving charge in the rest
frame of the model. Within the full far-field approximation we compute exact
expressions for the electric and magnetic fields, the angular distribution
of the power spectrum and the total emitted power in the m-th harmonic. We
also perform an expansion of the latter quantity in terms of the
electromagnetic LIV parameter and calculate the average degree of circular
polarization to first order in such a parameter. In both cases we find,
under adequate circumstances, the appearance of rather unexpected and large
amplifying factors, which go together with the otherwise negligible naive
expansion parameter. This opens up the possibility of selecting
astrophysical sources where these amplifying factors are important, to
explore further constraints imposed upon the LIV parameters by synchrotron
radiation measurements.
\end{abstract}

\pacs{11.30.Cp, 41.60.Ap, 03.50.Kk, 95.30.Gv}
\maketitle

\section{Introduction}

The possibility of exploring Lorentz invariance violations (LIV)
at the highest possible energies has been further motivated with
the proposal of Ref. \cite{AC98} that photons may acquire modified
energy-dependent dispersion relations (MDR). The effects on the
photon velocity could be substantially amplified when looking both
at cosmological and high energy sources (gamma ray burts, for
example) to allow for an observational detection of such minute
corrections. This idea has been further generalized to massive
spin 1/2 particles, in such a way that the typical form of such
modified dispersion relations is%
\begin{equation}
\omega^{2}(k)=k^{2}\;\pm\xi\frac{k^{3}}{M},\quad
E^{2}(p)=p^{2}+m^{2}+\eta_{R,L}\frac{p^{3}}{M},  \label{DR}
\end{equation}
respectively. Here $M$ is a scale which signals the onset of LIV and is
usually associated with the Planck mass $M_{P}$. This suggests that quantum
gravity could be responsible for such a breaking of Lorentz covariance. This
proposal has been further explored in Loop Quantum Gravity, where
constructions of increasing degree of sophistication have managed to produce
modified effective actions for photons and fermions encoding the relations (%
\ref{DR}). Basically, the starting point of these approaches has
been the well defined Hamiltonian operators of the quantum theory
together with an heuristical characterization of the semiclassical
ground state \cite{GP,URRU0,TH0,SMOLIN0}. String theory has also
provided a possible connection between quantum gravity and LIV
\cite{kostelecky1,ELLIS1}.

The MDR (\ref{DR}) generically arise from Lorentz-violating
dimension five operators in the Lagrangian, upon which we focus in
the following. On the phenomenological side the main emphasis has
been set on the determination of observational bounds for the LIV
parameters appearing in the MDR. A partial list of references is
given in \cite{analysis,JACOBSON}. Additionally, effective field
theories, not necessarily related to quantum gravity, have
been put forward to analyze the possible consequences of LIV \cite{MYPOS,POSP}%
. In this work we restrict ourselves to classical theories, and
therefore we do not address the fine-tuning problems arising from
quantum corrections associated with the coexistence of space
granularity and preferred reference frames \cite{FINETUN}.

Here we are concerned with the description of synchrotron
radiation coming from cosmological sources, which have already
been used to set bounds upon the parameters
$\xi,\eta_{L}\,,\eta_{R}$ in (\ref{DR}) \cite{JACOBSON}. Such
bounds are based on a set of very reasonable assumptions on how
some of the standard results of synchrotron radiation extend to
the Lorentz non-invariant situation. This certainly implies some
dynamical assumptions, besides the purely kinematical ones
embodied in (\ref{DR}). More specifically, we examine these
assumptions in the light of a particular model, which we choose to
be the classical version of the Myers-Pospelov effective theory
(MPET), which parameterizes LIV using dimension five
operators\cite{MYPOS}. Besides, a complete calculation of
synchrotron radiation in the context of this model is presented.
This constitutes an interesting problem on its own whose
resolution will subsequently allow the use of additional
observational information to put bounds upon the correction
parameters. We have in mind, for example, the polarization
measurements from cosmological sources. The case of gamma ray
bursts has recently become increasingly relevant\cite{CB},
although it is still at a controversial stage\cite{CONTR}. It
would also be interesting to compare the results from the MPET
with those arising from the Gambini-Pullin\cite{GP}, as well as
the Ellis et al.\cite{ELLIS1} electrodynamics. We postpone these
questions, together with most of the details in the MPET
calculation, for separate publications\cite{RALU}. Our
calculation rest heavily upon the work by Schwinger et al. on
synchrotron radiation, reported in Refs.
\cite{SCHWBOOK,SCH49,SCWANNP}. A partial list of previous studies
of electrodynamics incorporating LIV  via dimension three and
four operators is given in Ref. \cite{ALL}.

\section{Myers-Pospelov electrodynamics}

The Myers-Pospelov approach is based on an effective field theory that
describes Lorentz violations generated by dimension five operators
parameterized by the velocity $V^{\mu}$ of a preferred reference frame, not
taken as a dynamical field. These operators are assumed to be suppressed by
a large inverse mass factor, which we take to be $M_{P}^{-1}$, and hence can
be considered as perturbations at the classical level. In this framework we
analyze the electromagnetic radiation of a classical charged particle. To
avoid unnecessary complications we restrict the discussion to the case $%
V^{\mu}=\left( 1,\mathbf{0}\right) $, enough to show the main consequences
of the Lorentz violations. We take the unperturbed velocity of light $c$ in
vacuum equal to one.

The dynamics of a classical charged particle can be obtained from the
geometrical optics limit of a scalar charged field. In this case the
Myers-Pospelov action is%
\begin{equation}
S_{MP}=\int d^{4}x\;\left[ \partial_{\mu}\varphi^{\ast}\partial^{\mu}%
\varphi-\mu^{2}\varphi^{\ast}\varphi+i{\tilde{\eta}}\varphi^{\ast}\left(
V\cdot\partial\right) ^{3}\varphi\right] ,  \label{MPSCA}
\end{equation}
with $V\cdot\partial=V^{\mu}\,\partial_{\mu}$. In momentum space ($%
\varphi(x)=\varphi_{0}\;\exp i(p^{0}t-\mathbf{p\cdot x})$) and in the
reference frame where $V^{\alpha}=\left( 1,\mathbf{0}\right) $, the
dispersion relation becomes%
\begin{equation}
\left( p^{0}\right) ^{2}+{\tilde{\eta}}\left( p^{0}\right) ^{3}=\mathbf{p}%
^{2}+\mu^{2},  \label{EXDR}
\end{equation}
with ${\tilde{\eta}}=-\eta/M_{P}$, where $\eta$ is the dimensionless
constant employed in the parameterization of Jacobson et al.\cite{JACOBSON}.
The Eq. (\ref{EXDR}) is an exact relation in ${\tilde{\eta}}$.\ From here we
obtain the Hamiltonian for a massive particle, to second order in ${\tilde{%
\eta}}$
\begin{equation}
H({})=\left( \mathbf{p}^{2}+\mu^{2}\right) ^{1/2}-\frac{1}{2}{\tilde{\eta}}%
\left( \mathbf{p}^{2}+\mu^{2}\right) +\frac{5}{8}{\tilde{\eta}}^{2}\left(
\mathbf{p}^{2}+\mu^{2}\right) ^{3/2}+O({\tilde{\eta}}^{3}).  \label{H2OR}
\end{equation}
In what follows we consider the interaction of a particle of mass $\mu$ and
charge $q$ \ with a static magnetic field ($\phi=0,\;\mathbf{A}(x_{i})$) via
the standard minimal coupling. From the corresponding Hamiltonian equations
we obtain the acceleration
\begin{equation}
\mathbf{\ddot{r}}=\frac{q}{E}\left( 1-\frac{3}{2}{\tilde{\eta}}E+\frac{9}{4}{%
\tilde{\eta}}^{2}E^{2}\right) \left( \mathbf{v\times B}\right) ,  \label{acc}
\end{equation}
with $E=H(\mathbf{p}-q\mathbf{A})$ being the energy of the
particle. As in the usual case, this means that: (i) the magnitude
$|\mathbf{v|}$ is constant and (ii) the projection of the orbit in
a plane orthogonal to $\mathbf{B=\nabla\times A}$
is circular, with a Larmor frequency%
\begin{equation}
\omega_{0}=\frac{|q|B}{E}\left( 1-\frac{3}{2}{{\tilde{\eta}}}E+\frac{9}{4}{%
\tilde{\eta}}^{2}E^{2}\right) .
\end{equation}
Here we consider only the circular motion in a plane orthogonal to $\mathbf{B%
}$ and we take the $z$-axis in the direction of the constant magnetic field.
Using the standard definition\ $\beta=|\mathbf{v|}$, the solution to the
equations of motion can be written as
\begin{equation}
\mathbf{r(}t)\mathbf{=}\frac{\beta}{\omega_{0}}\left(
\cos\omega_{0}t,\sin\omega_{0}t,0\right) ,  \label{rt}
\end{equation}
which identifies the Larmor radius of the orbit as $R=\beta/\omega_{0}$. The
modified relation defining $\beta$ in terms of the particle energy is
\begin{equation}
1-\beta^{2}=\frac{\mu^{2}}{E^{2}}\left[ 1+2\frac{{\tilde{\eta}}E^{3}}{\mu
^{2}}-\frac{15}{4}\frac{{\tilde{\eta}}^{2}E^{4}}{\mu^{2}}+O({\tilde{\eta}}%
^{3})\right] .  \label{UMB2}
\end{equation}
Let us remark that we maintain the standard definition of the Lorentz factor
as $\gamma=\left( 1-\beta^{2}\right) ^{1/2}$. The current produced by the
corresponding charged particle is
\begin{equation}
\mathbf{j}(t,\mathbf{r})=q\delta^{3}(\mathbf{r}-\mathbf{r}(t))\,\mathbf{%
\dot {r}}(t),  \label{CURRENT}
\end{equation}
where $\mathbf{r}(t)$ is given by Eq.(\ref{rt}).

The Myers-Pospelov action for the electromagnetic field is%
\begin{equation}
S_{MP}=\int d^{4}x\left[ -\frac{1}{4}F_{\mu \nu }F^{\mu \nu }-4\pi \,j^{\mu
}A_{\mu }+{\tilde{\xi}}\left( V^{\alpha }F_{\alpha \delta }\right) (V\cdot
\partial )(V_{\beta }\tilde{F}^{\beta \delta })\right] ,  \label{mpa}
\end{equation}%
where ${\tilde{\xi}}=\xi /M_{P}$, with $\xi $ being a
dimensionless parameter. As usual we have $F_{\mu \nu }=\partial
_{\mu }A_{\nu }-\partial _{\nu }A_{\mu }$, $j^{\mu }=(\rho
,\mathbf{j})$ and we are using the conventions in Jackson
\cite{JACK}. Clearly this is a gauge theory and implies the
conservation of the electromagnetic current $j^{\mu}$. The
definition of $F_{\mu \nu }$ leads to the standard homogeneous
Maxwell equations. The source dependent equations of motion
obtained from the action (\ref{mpa}),
written in the rest frame $V^{\alpha }=(1,\mathbf{0})$, are%
\begin{equation}
\nabla \cdot \mathbf{E}=4\pi \rho ,\qquad -\frac{\partial \mathbf{E}}{%
\partial t}+\nabla \times \mathbf{B}+\tilde{\xi}\frac{\partial }{\partial t}%
\left( -\nabla \times \mathbf{E}+\frac{\partial \,\mathbf{B}}{\partial t}%
\right) =4\pi \mathbf{j}.  \label{MODMAX}
\end{equation}%
Introducing the standard vector potential field $A^{\mu }=\left( \phi ,%
\mathbf{A}\right) $ and using the radiation gauge $\nabla \cdot \mathbf{A}=0$
we are left with%
\begin{equation}
\frac{\partial ^{2}\mathbf{A}}{\partial t^{2}}-\nabla ^{2}\mathbf{A}+2\tilde{%
\xi}\nabla \times \frac{\partial ^{2}\mathbf{A}}{\partial t^{2}}=4\pi
\,\left( \mathbf{j-}\nabla \frac{1}{\nabla ^{2}}\nabla \cdot \mathbf{j}%
\right) \equiv 4\pi \mathbf{j}_{T}.  \label{ema}
\end{equation}%
Let us recall that in this gauge the electric and magnetic fields in the
radiation approximation reduce to%
\begin{equation}
\mathbf{E}=-\frac{\partial \mathbf{A}}{\partial t},\;\;\;\;\mathbf{B}=\nabla
\times \mathbf{A}.  \label{RADF}
\end{equation}%
The energy momentum tensor $T_{\mu \nu }$ for this modified electrodynamics
is given by%
\begin{equation}
T_{\;0}^{0}=\frac{1}{8\pi }(\mathbf{E}^{2}+\mathbf{B}^{2})-\frac{\tilde{\xi}%
}{4\pi }\mathbf{E\,\cdot \,}\frac{\partial \mathbf{B}}{\partial t},\qquad
\mathbf{S}\mathbf{=}\frac{1}{4\pi }\mathbf{E}\times \mathbf{B}-\frac{\tilde{%
\xi}}{4\pi }\mathbf{E}\times \frac{\partial \mathbf{E}}{\partial t},
\label{pv}
\end{equation}%
which satisfies the usual conservation equation $\partial _{t}T^{00}+\mathbf{%
\nabla }\cdot \mathbf{S}=0$ outside the sources. These expressions are exact
in $\tilde{\xi}$.

To solve the equation of motion for $\mathbf{A}$ it is convenient to work in
momentum space, where Eq. (\ref{ema}) becomes%
\begin{equation}
\left( -\omega ^{2}+k^{2}-2i\tilde{\xi}\omega ^{2}\,\mathbf{k}\,\times
\right) \mathbf{A(}\omega \mathbf{,k)}=4\pi \,\mathbf{j}_{T}\mathbf{(}\omega
,\mathbf{k)},  \label{EQAMOM}
\end{equation}%
with $\mathbf{j}_{T}\mathbf{(}\omega ,\mathbf{k)=j-\hat{k}}\left( \mathbf{%
\hat{k}\cdot j}\right) $ and  $\mathbf{\hat{k}}=\mathbf{k}/k$ with
$k=|\mathbf{k}|$.  The notation we employ for the different types
of Fourier transforms is via the corresponding arguments. For
example, if  $F(t, \mathbf{r)}$ denotes the function in
space-time, $F(\omega, \mathbf{r)}$ denotes the Fourier
transformed function to frequency space, while $F(\omega,
\mathbf{k})$
denotes the Fourier transformed function to frequency and momentum spaces. Equation (\ref%
{EQAMOM}) can be diagonalized using a circular polarization basis
yielding radiation fields of definite helicity
\begin{equation}
\mathbf{A}^{\lambda }\mathbf{(}\omega \mathbf{,k)}=\frac{1}{2}\left[ \mathbf{%
A}-\left( \mathbf{\hat{k}}\cdot \mathbf{A}\right) \mathbf{\hat{k}}+i\lambda
\mathbf{\hat{k}}\times \mathbf{A}\right] ,
\end{equation}%
with $\lambda =\pm 1$, leading to
\begin{equation}
\left( -\omega ^{2}+k^{2}-\lambda 2\tilde{\xi}\omega ^{2}k\right) \mathbf{A}%
^{\lambda }\mathbf{(}\omega \mathbf{,k)}=4\pi \mathbf{j}_{T}^{\lambda }%
\mathbf{(}\omega \mathbf{,k)},  \label{DEFPOL}
\end{equation}%
where%
\begin{equation}
\mathbf{j}_{T}^{\lambda }\mathbf{(}\omega
\mathbf{,k)}=\frac{1}{2}\left(
\mathbf{j}_{T}+i\lambda \mathbf{\hat{k}}\times \mathbf{j}_{T}\right) =\frac{1%
}{2}\left[ \mathbf{j-}\left( \mathbf{\hat{k}\cdot j}\right) \mathbf{\hat{k}}%
+i\lambda \mathbf{\hat{k}}\times \mathbf{j}\right]=\mathbf{j}^{\lambda }%
\mathbf{(}\omega \mathbf{,k)}.  \label{POLCURR}
\end{equation}%

The Eqs. (\ref{DEFPOL}) imply the following dispersion relations
for the propagating fields
\begin{equation}
\omega _{\lambda }^{2}=\frac{k^{2}}{1+2\lambda \tilde{\xi}k}.
\end{equation}%
To first order in $\tilde{\xi}$\textbf{,}$\;$ they have the same
form as those corresponding to the photon in Eq. (\ref{DR}).

Equation (\ref{DEFPOL}) can be solved in terms of the retarded
Green  functions with definite helicity
\begin{equation}
G_{ret}^{\lambda}(\omega, \mathbf{k})=\left[ \frac{1}{\left(
-\omega ^{2}+k^{2}-\lambda 2\tilde{\xi}\omega ^{2}k\right)
}\right] _{\omega \rightarrow \omega +i\epsilon }.
\end{equation}
It is convenient to calculate%
\begin{equation}
G_{ret}^{\lambda }\mathbf{(}\omega \mathbf{,r-r}^{\prime }\mathbf{)=}\int
\frac{d^{3}\mathbf{k}}{\left( 2\pi \right) ^{3}}e^{i\mathbf{k}\cdot (\mathbf{%
r-r}^{\prime })}G_{ret}^{\lambda }\mathbf{(}\omega \mathbf{,k)=}\frac{1}{%
4\pi R}\frac{n(\lambda z)}{\sqrt{1+z^{2}}}e^{in(\lambda z)\omega R}
\label{GRETGF}
\end{equation}%
where $R=|\mathbf{r}-\mathbf{r}^{\prime }|$. Here we introduce
the polarization-dependent refraction index $n(\lambda z)$
\begin{equation}
n(\lambda z)=\sqrt{1+z^{2}}+\lambda z,\;\;\mathbf{\ }z=\tilde{\xi}\omega .
\label{REFIND}
\end{equation}%
In this way, the fields $\bf A^{\lambda}$ in Eq. (\ref{DEFPOL})
have well defined phase velocities $v_{\lambda }=1/n(\lambda z)$
and the situation can be described as the propagation of photons
in a dispersive birefringent medium.

The Green functions (\ref{GRETGF}) determine the corresponding
potentials in the far field approximation with the standard replacements $%
1/R\simeq 1/|\mathbf{r}|\equiv 1/r$ in the denominator and $R\simeq r-%
\mathbf{\hat{n}}\cdot \mathbf{r}^{\prime }$ in the phase, with $\mathbf{\hat{%
n}}=\mathbf{r}/r$ been the direction of observation. Using
(\ref{GRETGF}) we obtain
\begin{equation}
\mathbf{A}^{\lambda }(\omega ,\mathbf{r}) = 4\pi \int d^{3}\mathbf{r}%
^{\prime }\;G_{ret}^{\lambda }\mathbf{(}\omega \mathbf{,r-r}^{\prime }%
\mathbf{)\;j}_{T}^{\lambda }\mathbf{(}\omega \mathbf{,\mathbf{r}^{\prime })=}%
\frac{1}{R}\frac{n(\lambda z)}{\sqrt{1+z^{2}}}e^{in(\lambda
z)\omega r}\int
d^{3}\mathbf{r}^{\prime }\;e^{-i\left[ n(\lambda z)\omega \mathbf{\hat{n}}%
\right] \cdot \mathbf{r}^{\prime }}\mathbf{\;j}_{T}^{\lambda }\mathbf{(}%
\omega \mathbf{,\mathbf{r}^{\prime })},
\end{equation}
and thus we finally get
\begin{equation}
\mathbf{A}^{\lambda }(\omega ,\mathbf{r}) =\frac{1}{r}\frac{%
n(\lambda {z})}{\sqrt{1+{z^{2}}}}e^{in(\lambda {z})\omega r}\mathbf{j}%
^{\lambda }(\omega ,\mathbf{{k}_{\lambda }}), \label{APOL}
\end{equation}%
in the radiation approximation. Let us emphasize that  the
momenta $\bf {k}_{\lambda }=n(\lambda {z})\omega \mathbf{\hat{n}}$
{are} fixed in terms of  the frequency and the direction of
observation. As usual, the integration over
$d^{3}\mathbf{r}^{\prime }$  has been
conveniently written as  the Fourier transform  in momentum space $\mathbf{j}%
_{T}^{\lambda }(\omega ,\mathbf{{k}_{\lambda }})$ of  the function $\mathbf{j}%
_{T}^{\lambda }\mathbf{(}\omega \mathbf{,\mathbf{r}^{\prime })}$
in coordinate space.  In the last term of (\ref{APOL}) we have
used the relation (\ref {POLCURR}). The full vector potential is
given by the superposition
\begin{equation}
\mathbf{A}(\omega ,\mathbf{r})=\frac{1}{r}\frac{1}{\sqrt{1+{z^{2}}}}%
\sum_{\lambda =\pm 1}n(\lambda {z})e^{in(\lambda {z})\omega r}\mathbf{j}%
^{\lambda }(\omega ,\mathbf{{k}_{\lambda }}).  \label{VECPOT}
\end{equation}
The magnetic and electric fields satisfy
\begin{equation}
\mathbf{B}(\omega ,\mathbf{r})=\sqrt{1+{z^{2}}}\,\mathbf{\hat{n}\times E}%
(\omega ,\mathbf{r})-i{z}\,\mathbf{E}(\omega ,\mathbf{r}).  \label{BAFE}
\end{equation}%
At this point it is interesting to observe that the above relation, together
with the expression for the Poynting vector in (\ref{pv}), yields the $z$%
-exact, manifestly positive-definite result%
\begin{equation}
\mathbf{S}=\frac{1}{4\pi }\sqrt{1+z^{2}}\ \mathbf{E}^{\ast }(\omega )\cdot
\mathbf{E}(\omega )=\frac{1}{4\pi }\left( \frac{n(z)+n(-z)}{2}\right) \
\mathbf{E}^{\ast }(\omega )\cdot \mathbf{E}(\omega ).  \label{SIMPV}
\end{equation}%
This reproduces the standard expression for the Poynting vector in a medium
with refraction index $n$ by the substitution $(n(z)+n(-z))/2\rightarrow n$.

The use of  Eqs.(\ref{RADF}) and (\ref{VECPOT}) leads to the
following electric
and magnetic fields in the radiation approximation%
\begin{align}
\mathbf{B(}\omega ,\mathbf{r})& =\frac{1}{r}\frac{\omega }{\sqrt{1+z^{2}}}%
\sum_{\lambda =\pm 1}\lambda \,n^{2}(\lambda z)e^{in(\lambda z)\omega r}%
\mathbf{j}^{\lambda }(\omega ,\mathbf{k}_{\lambda }),  \label{BDIELD} \\
\mathbf{E(}\omega ,\mathbf{r})& =\frac{1}{r}\frac{i\,\omega }{\sqrt{1+z^{2}}}%
\sum_{\lambda =\pm 1}n(\lambda z)e^{in(\lambda z)\omega r}\mathbf{j}%
^{\lambda }(\omega ,\mathbf{k}_{\lambda }).  \label{EFIELD}
\end{align}%
Finally, using for example the methods in Ref. \cite{SCHWBOOK}, we arrive at
the following general expression for the angular distribution of the power
spectrum in the MPET
\begin{equation}
\frac{d^{2}P(T)}{d\omega d\Omega }=\frac{1}{4\pi ^{2}}\frac{\omega ^{2}}{%
\sqrt{1+z^{2}}}\int_{-\infty }^{\infty }d\tau \ e^{-i\omega \tau
}\,\sum_{\lambda =\pm 1}n^{2}(\lambda z)j_{k}^{\ast }\left( T+\tau /2,%
\mathbf{k}_{\lambda }\right) \, P_{kr}^{\lambda } \, j_{r}\left( T-\tau /2,%
\mathbf{k}_{\lambda }\right)   \label{poo}
\end{equation}
The projector $P_{kr}^{\lambda }$ can be read from the relation
among the components of the polarized and total currents given
in  Eq. (\ref{POLCURR}). The total current is written here in the
representation $j_r(t, \mathbf{k})$.

\section{Synchrotron Radiation}

We now consider the radiation of a charged particle moving in a
circular orbit, in a plane perpendicular to a constant magnetic
field $\mathbf{B}$ parallel to the $z$-axis, according to
Eq.(\ref{rt}). The direction of observation is given by
$\mathbf{\hat{n}}=\mathbf{r}/r=(\sin\theta,0,\cos\theta)$ and the
products of currents that appear in Eq.(\ref{poo}) lead us to
consider the
following combinations of the velocity ${\mathbf v}(t)$=$ \mathbf{%
\dot {r}}(t)$,  according to  Expression (\ref{CURRENT}),
\begin{eqnarray}
\mathbf{v}(T+\tau /2)\cdot \mathbf{v}(T-\tau /2) &=&q^2\beta
^{2}\cos \omega
_{0}\tau , \\
\mathbf{\hat n\cdot v}(T+\tau /2)\times \mathbf{v}(T-\tau /2)
&=&-q^2\beta ^{2}\cos \theta \sin \omega _{0}\tau .
\end{eqnarray}
The angular distribution of the radiated power spectrum is given
by the average over the macroscopic time $T$ of Eq.(\ref{poo}). As
in the usual case, it is convenient to write the relevant
quantities as a sum over the contribution of the harmonics
$m=0,1,2,\dots$. By doing this we have
\begin{equation}
\left\langle \frac{d^{2}P(T)}{d\omega d\Omega}\right\rangle =\sum
_{\lambda=\pm 1 }\,\sum_{m=0}^{\infty}\delta({\omega}-\omega_{m})\frac {%
dP_{m,\,\lambda}}{d\Omega},  \label{HARMEXP}
\end{equation}
with $\omega_{m}=m\omega_{0}$, $z_{m}=\tilde{\xi}\omega_{m}$ and%
\begin{equation}
\frac{dP_{m,\,\lambda}}{d\Omega}=\frac{\omega_{0}^{2}q^{2}}{4\pi}\frac {1}{%
\sqrt{1+z_{m}^{2}}}\left[ \lambda m\beta n(\lambda z_{m})J_{m}^{\,\prime
}(W_{\lambda m})+\,m\cot\theta\,J_{m}(W_{\lambda m})\right] ^{2}.
\label{ADMH}
\end{equation}
Here $\left\langle \;\right\rangle $ denotes the average over $T$, and $%
J_{m} $, $J_{m}^{\,\prime}$ are the Bessel functions of order $m$
and their derivatives respectively. The argument of the Bessel
functions is
\begin{equation}
W_{\lambda m}=m\,n(\lambda z_{m})\beta\sin\theta.  \label{WLM}
\end{equation}
Let us remark that the unpolarized angular distribution of the
power obtained from Eq. (\ref{ADMH}) coincides with the standard
result given in Eq. (38.42) of Ref.\cite{SCHWBOOK} in the limit
${\tilde{\xi}}=0$. From (\ref{ADMH}) we obtain that the angular
distribution of the averaged power  is anisotropic to first order
in $\tilde{\xi}$. We also calculate the total
averaged and integrated power radiated into the $m$-th harmonic%
\begin{align}
P_{m} & =\frac{q^{2}m\omega_{0}}{2R\,\sqrt{1+z_{m}^{2}}}\,\sum_{\lambda=\pm
1 }n(\lambda z_{m})\left[ 2\beta^{2}J_{2m}^{\,\prime}(2m\;n(\lambda
z_{m})\beta)\right.  \notag \\
& \left. -\left[ n^{-2}(\lambda z_{m})-\beta^{2}\right] \int
_{0}^{2mn(\lambda z_{m})\beta}dx\;J_{2m}(x)\right] ,  \label{PMDO}
\end{align}
which clearly indicates the contribution of each polarization. The above
result is exact in $z_{m}$ and the parity-violating contribution vanishes
after the angular integration.

Additional information regarding the LIV parameter $\tilde{\xi}$ can be
obtained from polarization measurements of the incoming radiation. To this
end we have calculated the exact expression for the $m$-th contribution to
the electric field in the radiation approximation $\mathbf{E}_{m}^{\lambda }=%
\mathbf{E}_{m}^{\lambda}(\omega_{m},\mathbf{k}_{\lambda})$%
\begin{equation}
\mathbf{E}_{m}^{\lambda}=-\sqrt{2}\pi(-i)^{m}\frac{q\beta}{r}\frac{\omega
_{m}\,n(\lambda z_{m})}{\sqrt{1+z_{m}^{2}}}e^{in(\lambda
z_{m})\omega_{m}r}\left( \beta\frac{dJ_{m}(W_{\lambda m})}{dW_{\lambda m}}%
+\lambda\frac {J_{m}(W_{\lambda m})}{\tan\theta}\;\right) \mathbf{e}%
_{\lambda},  \label{CPOLEF}
\end{equation}
where $\mathbf{e}_{\lambda}$ denotes the circularly polarized basis
\begin{equation}
\mathbf{e}_{\pm}=\frac{1}{\sqrt{2}}\left( \mathbf{e}_{\parallel}\pm i\mathbf{%
e}_{\perp}\right) ,\quad\mathbf{e}_{\parallel}=(0,1,0),\quad \mathbf{e}%
_{\perp}=(-\cos\theta,0,\sin\theta),  \label{CBASIS}
\end{equation}
in the chosen reference frame. The expression (\ref{CPOLEF}) allows for the
calculation of the corresponding Stokes parameters.

There are two ways in which the photon LIV parameter $\tilde{\xi}$ enters
the modified expressions for the physical quantities: (i) one is via
over-all multiplicative factors such as $1/\sqrt{1+z^{2}}$ or $n(\lambda z)$%
, where the dependence upon $\lambda\tilde{\xi}\omega$ makes these
contributions negligible for the range of observed frequencies. (ii) the
second possibility is through the dependence upon the variable $W_{\lambda
m} $ in the arguments of the Bessel functions. We have found that this case
provides some possibilities of producing observable effects because of the
appearance of additional amplifying factors, as we show below. Now, the
contributions to case (ii) are governed by the far field expansion of the
Green function phase%
\begin{equation}
n(\lambda{z})\omega\left\vert \mathbf{r}-\mathbf{r}^{\prime}\right\vert
\simeq\omega r\left( 1-\frac{\mathbf{n}\cdot\mathbf{r}^{\prime}}{r}+\lambda%
\tilde{\xi}\omega-\lambda\tilde{\xi}\omega\frac{\mathbf{n}\cdot\mathbf{r}%
^{\prime}}{r}+\frac{1}{2}\left( \frac{r^{\prime}}{r}\right) ^{2}\right) ,
\label{PHASE}
\end{equation}
where the term proportional to $\left(r^{\prime}/r\right)^{2}$ is always
neglected in the radiation approximation. Consistency demands that one had
better make sure that terms proportional to the LIV parameter $\tilde{\xi}$
are larger that the neglected one in order to properly include them in the
argument of the Bessel functions. To this end, in Table \ref{tab:a} we make
a rough estimation of the relevant parameters corresponding to some observed
cosmological objects. The notation is: $r\,[l.y.]$ is the distance of the
object to the earth, $\gamma$ is the Lorentz factor of the charged
particles, $B\,[Gauss]$ is the average magnetic field producing the
synchrotron radiation, $\omega\,[GeV]$ is the maximum observed frequency, $%
\omega _{0}\,[GeV]$ is the Larmor frequency and $r^{\prime}=R$ is an
estimation of the size of the radiating source, where $R$ is the Larmor
radius. Our general results are presented in the full far-field
approximation.

\begin{table}[ptb]
\caption{Typical parameters of some relevant astrophysical objects}
\label{tab:a}%
\begin{ruledtabular}
\begin{tabular}{cccccccccccc}
&$r$&$\gamma$&$B$& $\omega$
&$\omega_0$& $m$ &$\frac{m}{\gamma}$&$\frac{r'}{r}
$&$\frac{\omega}{M_P}$&$\frac{r'}{r}\frac{\omega}{M_P}$
&$\left(\frac{r'}{r}\right)^2$ \\
\hline CRAB& $10^4$ & $10^9$ &$10^{-3}$  &$10^{-1}$
& $10^{-30}$ &$10^{29}$& $10^{20}$ &$10^{-6}$&$10^{-20}$& $10^{-26}$&$10^{-12}$ \\
MARKARIAN& $10^{8}$ &$10^{11}$  &$10^{2}$  &$10^{4}$
& $10^{-26}$ & $10^{30}$ &$10^{19}$ &$10^{-14}$&$10^{-15}$&$10^{-29}$& $10^{-28}$ \\
GRB021206&$10^{10}$ & $10^{5}$ &$10^{4}$  &$10^{-3}$ & $10^{-18}$&
$10^{15}$&$10^{10}$
&$10^{-24}$&$10^{-22}$&$10^{-46}$&$10^{-48}$\\
\end{tabular}
\end{ruledtabular}
\end{table}
As indicated in Table \ref{tab:a}, the radiation of interest is dominated by
very high harmonics $m>>1$. Thus, it is convenient to use the large $m$
approximation for functions of the type $J_{m}(mx)$ together with their
derivatives \cite{SCH49}. In the regime $1-n^{2}\beta^{2}>0$ we find%
\begin{align}
P_{\lambda m} & =\frac{q^{2}m\omega_{0}}{\sqrt{3}\pi R}\frac{1}{%
1+n^{2}(\lambda{z_{m}})}\left\{ \left( \frac{3}{2{\tilde{m}}_{\lambda c}}%
\right) ^{2/3}\int_{m/\tilde{m}_{\lambda c}}^{\infty}dxK_{5/3}\left(
x\right) \right.  \notag \\
& \left. -2\left( \frac{3}{2{\tilde{m}}_{\lambda c}}\right)
^{4/3}\,K_{2/3}\left( \frac{m}{\tilde{m}_{\lambda c}}\right) \right\} ,
\label{HMPO}
\end{align}
which introduces the cut-off harmonic%
\begin{equation}
\tilde{m}_{\lambda c}=\frac{3}{2}\left[ 1-\beta^{2}\,n^{2}(\lambda{z_{m}})%
\right] ^{-3/2},  \label{HCOFF}
\end{equation}
such that for $\,m>\tilde{m}_{\lambda c}$ the power decreases as
$P_{\lambda m}\approx e^{-m/\tilde{m}_{\lambda c}}$. This is
basically the result of Ref. \cite{SCWANNP} extended to polarized
radiation. Here $K_{p/q}$ denote the Bessel functions of
fractional index. Within the same large-$m$ approximation, the
integrated power in the $m$-th harmonic can be expanded
to second order in ${\tilde{\xi}}$ yielding%
\begin{equation}
P_{m}=\frac{q^{2}\omega}{\sqrt{3}\pi R\gamma^{2}}\left\{ \frac{m_{c}}{m}{%
\kappa}\left( \frac{m}{m_{c}}\right) -\frac{2}{\gamma^{2}}K_{2/3}\left(
\frac{m}{m_{c}}\right) +2\left( \frac{{\tilde{\xi}}\,m\omega\beta}{\gamma}%
\right) ^{2}K_{2/3}\left( \frac{m}{m_{c}}\right) \right\} ,  \label{PMEXP}
\end{equation}
where $m_{c}=3\gamma^{3}/2$ and ${\kappa}(x)=x\int_x^{\infty}dy\,K_{5/3}(y)$
is the bremstrahlung function \cite{ERBER}. Let us emphasize the appearance
of the combination ${\tilde{\xi }}\,\omega
m/\gamma=\xi(\omega/M_{P})(m/\gamma)$ as the expansion parameter in (\ref%
{PMEXP}). From Table \ref{tab:a} we see that this combination is not
necessarily a small number, which signals the possibility that such
corrections might be relevant in setting bounds upon $\tilde{\xi}$. This
rather unexpected effect is due to the amplifying factor $\left(
m/\gamma\right)$.

Another possibility for observable effects due to ${\tilde{\xi}}$ is to look
at the averaged degree of circular polarization%
\begin{equation}
\Pi_{\odot}=\frac{\left\langle P_{+}(\omega)-P_{-}(\omega)\right\rangle }{%
\left\langle P_{+}(\omega)+P_{-}(\omega)\right\rangle },  \label{CIRCPOL}
\end{equation}
where $P_{\lambda}(\omega)$ is the total power distribution per unit
frequency and polarization $\lambda$, so that $P_{\lambda}(\omega)=P_{m%
\lambda}/\omega_{0}$. The average here is calculated with respect to an
energy distribution of the relativistic electrons, which we take to be $%
N(E)dE=CE^{-p}dE$, in some energy range $E_{1}<E<E_{2}$. A standard choice
for the exponent is $2<p<3$. The result is%
\begin{equation}
\Pi_{\odot}={\tilde{\xi}}\omega\left( \frac{\mu\omega}{qB}\right) \,\Pi(p),
\label{CPEXP}
\end{equation}
where $\Pi(p)$ will be estimated in what follows. Again, we have the
presence of an amplifying factor in Eq. (\ref{CPEXP}), given by $%
(\mu\omega/qB)$, which is independent of the form of $\Pi(p)$ and not
necessarily a small number. An estimation of this factor in the zeroth-order
approximation (${\tilde \xi}=0={\tilde \eta}$), which is appropriate in (\ref%
{CPEXP}), yields $(\mu\omega/q B )=\omega/(\omega_0
\gamma)=m/\gamma$. The expression (\ref{CPEXP}) is analogous to
the well-known average of the degree of linear polarization
$\Pi_{LIN}=(p+1)/(p+7/3)$, under the same energy distribution for
the electrons \cite{RL}.

Now let us compute the degree of circular polarization (\ref{CIRCPOL}) in
order to verify our previous statement. To this end we start from Eq. (\ref%
{HMPO}) and set $\beta=1=n(\lambda z_{m})$ everywhere, except in the
critical terms involving ${\tilde{m}}_{\lambda c}$ from where the
corrections arise. We should also take into account that most of the
radiation arises from the terms with $m\approx{\tilde{m}}_{\lambda c}>>1$,
where $K_{2/3}(1)=0.49,\,\,{\kappa}(1)=0.65$. Thus the dominant term is%
\begin{equation}
P_{\lambda}(\omega)=D\,\left[ \left( \frac{1}{\tilde{m}_{\lambda c}}\right)
^{2/3}\frac{\tilde{m}_{\lambda c}}{m}{\kappa}(m/\tilde{m}_{\lambda c})\right]
,  \label{DOMTERM}
\end{equation}
where $D$ is a constant. Starting from the definition (\ref{CIRCPOL}),
making an expansion in $u=\lambda{\tilde{\xi}}\omega$ and recalling that $%
\tilde
{m}_{\lambda c}=\tilde{m}_{\lambda c}(u)$, we find%
\begin{equation}
\Pi_{\odot}={\tilde{\xi}}\omega\frac{\left\langle \left[ \frac{dP_{\lambda
}(\omega)}{du}\right] _{{\tilde{\xi}}=0,\,\,{\tilde{\eta}}=0}\right\rangle }{%
\left\langle \left[ P_{\lambda}(\omega)\right] _{{\tilde{\xi}}=0,\,\,{\tilde{%
\eta}}=0}\right\rangle }+O({{\tilde{\xi}}}^{2},\;{\tilde{\xi}}{\tilde{\eta}}%
,\dots).  \label{CIRCPOL1}
\end{equation}
Since $n(u)=\sqrt{1+u^{2}}+u$ we can rewrite $d/du$ in terms of $d/dn$.
Besides, $\tilde{m}_{\lambda c}$ depends only on the combination $n\beta$ so
that we can further go to $d/d\beta$ obtaining%
\begin{equation}
\left[ \frac{dP_{\lambda}(\omega)}{du}\right] _{{\tilde{\xi}}=0,\,\,{\tilde {%
\eta}}=0}=D\left[ \frac{2n^{2}}{1+n^{2}}\frac{\beta}{n}\right] _{{{\tilde{\xi%
}}=0,\,\,{\tilde{\eta}}=0}}\,\frac{d}{d\beta}\left[ \left( \frac{1}{m_{c}}%
\right) ^{2/3}\frac{m_{c}}{m}{\kappa}(m/m_{c})\right] ,  \label{DER1}
\end{equation}
where the first parenthesis on the RHS gives $1$ upon evaluation, and we
have already taken the limit $n=1,{\tilde{\eta}}=0$ in the second
parenthesis. In this way we now have%
\begin{equation}
m_{c}=\frac{3}{2}\gamma^{3},\qquad E=\mu\gamma.  \label{KAPCER}
\end{equation}
Next, we consider the variable%
\begin{equation}
x=\frac{m}{m_{c}}=\frac{2}{3}\frac{\omega}{\omega_{0}}\gamma^{-3}=A^{2}\,%
\gamma^{-2},\quad A^{2}=\frac{2}{3}\left( \frac{\mu\omega}{qB}\right) ,
\label{XVAR}
\end{equation}
where $A^{2}$ is proportional to the amplifying factor in (\ref{CPEXP}). The
above leads to $\gamma=A\,x^{-1/2}$. Again, a successive change of
independent variables yields%
\begin{equation}
d/d\beta=\beta\,\gamma^{3}\,d/d\gamma=-2A^{2}\,d/dx,  \label{CHARUL}
\end{equation}
where we have set $\beta=1$ in the corresponding factor. Finally, we have to
analyze the term $\left( x\,m_{c}^{2/3}\right)$ that multiplies\textbf{\ } $%
\kappa(x)$ inside the square bracket of Eqs. (\ref{DOMTERM}) and (\ref{DER1}%
). Substituting $m_{c}=3\gamma^{3}/2=3\,A^{3}\,x^{-3/2}/2$ we find that $%
\left( x\,m_{c}^{2/3}\right) =(3A^{3}/2)^{2/3}$ is a constant that cancels
when taking the ratio in (\ref{CIRCPOL1}). Thus we are left with%
\begin{equation}
\Pi_{\odot}={\tilde{\xi}}\omega\;\left( \frac{\mu\omega}{qB}\right) \,\left(
-\frac{4}{3}\right) \,\frac{\left\langle \frac{d\kappa(x)}{dx}\right\rangle
}{\left\langle \kappa(x)\right\rangle },  \label{CPEXP1}
\end{equation}
where the average $\left\langle .\right\rangle $ over $E$ has been replaced
by one over $x$, via the change of variables $E=\mu\,A\,x^{-1/2}$, with all
the constant factors cancelling in the ratio of the two integrals. It should
be pointed out that the factor $A^{2}$ arising from the derivative in the
numerator and responsible for the amplifying factor survives after taking
this ratio. For the purpose of making an estimation of $\Pi(p)$ we take the
energy range to be $\,0\leq E\leq\infty$. We then obtain
\begin{equation}
\Pi_{\odot}=-\frac{4}{3}{\tilde{\xi}}\omega\;\left( \frac{\mu\omega}{qB}%
\right) \,\,\frac{\int_{0}^{\infty}x^{\left( p-3\right) /2}\frac{d{\kappa}(x)%
}{dx}\;dx}{\int_{0}^{\infty}x^{\left( p-3\right) /2}\;{\kappa}(x)\;dx}.
\label{CPEXP2}
\end{equation}
Using the expresion\cite{RL}%
\begin{equation}
\int_{0}^{\infty}x^{\mu}\;{\kappa}(x)\;dx=\frac{2^{\mu+1}}{\mu+2}%
\Gamma\left( \frac{\mu}{2}+\frac{7}{3}\right) \Gamma\left( \frac{\mu}{2}+%
\frac{2}{3}\right) ,\qquad\mu+1/3>-1,
\end{equation}
and comparing with Eq. (\ref{CPEXP}), we finally get%
\begin{equation}
\Pi(p)=\frac{(p-3)\left( 3p-1\right) }{3\left( 3p-7\right) }\,\frac {(p+1)}{%
(p-1)}\frac{\Gamma\left( \frac{p}{4}+\frac{13}{12}\right) \Gamma\left( \frac{%
p}{4}+\frac{5}{12}\right) }{\Gamma\left( \frac{p}{4}+\frac{19}{12}\right)
\Gamma\left( \frac{p}{4}+\frac{11}{12}\right) },\qquad p>7/3  \label{PIP}
\end{equation}
The constraint $p>7/3$ is required to avoid the divergence of the integral
in the numerator of Eq. (\ref{CPEXP2}) at $x=0$. On more realistic grounds,
one should avoid the infinite upper limit of $E$, (the zero lower limit of $%
x $). In this case the divergence of the integral at $x=0,$ and hence the
mathematical constraint on $p$, disappears, but the expression for $\Pi(p)$
becomes more complicated. In any case, the most important feature of the
result in Eqs. (\ref{CPEXP}) and (\ref{PIP}) is the presence of the
amplifying factor $(\mu\omega/qB)$, which is independent of such details.

\section{Summary and Outlook}

We have presented a summary of a complete calculation of synchrotron
radiation in the Myers-Pospelov effective model. The results can be
understood in terms of the standard electrodynamics description of a
parity-violating birrefringent media with a given helicity-dependent
refraction index $n(\lambda{\tilde{\xi}}\omega)$ containing corrections
arising from the electromagnetic sector of the theory. For each
polarization-dependent refraction index, the results of Ref. \cite{SCWANNP}
are recovered. In particular we obtain that the opening angle of each
contribution to the polarized radiation is given by%
\begin{equation}
\delta\theta\approx m^{-1/3}\approx m_{\lambda
c}^{-1/3}=[1-\beta^{2}(E)n^{2}(\lambda{\tilde{\xi}}\omega)]^{1/2}.
\label{OPANG}
\end{equation}
Corrections also arise from the charged particle sector and are encoded in
the modified dependence of the $\beta$ factor upon the particle energy,
according to Eq. (\ref{UMB2}).

In the case of the CRAB nebula, Table {\ref{tab:a}} tells us that ${\tilde {%
\xi}}\omega=\xi\omega/M_{P}<<(r^{\prime}/r)^{2}$, even for the upper bound
of $\xi\approx10^{-4}$ given by Gleiser and Kozameh in Ref. {\cite{analysis}}%
. In this way, consistency requires that the phase of the Green function (%
\ref{PHASE}) reduces to $\omega\left( r-\mathbf{\hat{n}}\cdot \mathbf{%
r^{\prime}}\right)$. This is equivalent to set ${\tilde{\xi}}=0,n=1$ in all
the arguments of the Bessel functions. This leads to
\begin{equation}
\delta\theta\approx{\gamma}^{-1}(E),\qquad\omega_{c}=\frac{qB}{E}\gamma
^{3}(E),
\end{equation}
where $\gamma(E)$ still includes corrections depending upon the fermionic
parameter ${\tilde{\eta}}$ according to Eq. (\ref{UMB2}). In others words,
the corresponding results in Ref. \cite{JACOBSON} are recovered for this
particular situation.

It would be interesting to search for astrophysical objects in which the
amplifying factors are important, focusing on the observation of the average
circular polarization, of possible corrections to the average linear
polarization and of the Stokes parameters, among other possibilities.

\section*{Acknowledgments}

LFU acknowledges partial support from the projects CONACYT-40745F and
DAGAPA-IN104503-3. RM acknowledges partial support form CONICET-Argentina,
and kind hospitality and support from the Instituto de Ciencias Nucleares,
UNAM, where most of this work was developed.

\end{document}